# Influence of high pressure on Ce$^{3+}$ luminescence in LuAlO$_3$ and YAlO$_3$ single crystals and single crystalline layers

*Lev-Ivan Bulyk [a], Ajeesh Kumar Somakumar [a], Hanka Przybylińska [a], P. Ciepielewski [a,h],*

*Yu. Zorenko [b], Ya. Zhydachevskyy [a], I. Kudryavtseva [c], V. Gorbenko [b], A. Lushchik [c],*

*M. G. Brik [c, d, e, f], Y. Syrotych [b], S. Witkiewicz-Łukaszek [b], A. Fedorov [g] and Andrzej Suchocki [a, *]*

[a] Institute of Physics, Polish Academy of Sciences, Al. Lotników 32/46, 02-668, Warsaw, Poland

[b] Institute of Physics, Kazimierz Wielki University in Bydgoszcz, Powstańców Wielkopolskich str., 2, 85-090 Bydgoszcz, Poland.

[c] Institute of Physics, University of Tartu, W. Ostwaldi 1, 50411 Tartu, Estonia

[d] College of Sciences & CQUPT-BUL Innovation Institute, Chongqing University of Posts and Telecommunications, Chongqing 400065, People's Republic of China

[e] Faculty of Science and Technology, Jan Dlugosz University, Armii Krajowej 13/15, PL-42200 Czestochowa, Poland

[f] Academy of Romanian Scientists, Ilfov Street, no 3, 050044, Bucharest, Romania

[g] SSI Institute for Single Crystals, National Academy of Sciences of Ukraine, 61178 Kharkiv, Ukraine

[h] Lukasiewicz Res Network, Inst Microelect & Photon, Al Lotnikow 32-46 Str, PL-02668 Warsaw, Poland

*corresponding author

## Abstract

Results of spectroscopic studies at ambient and high pressures of a LuAlO$_3$:Ce$^{3+}$ (LuAP:Ce) single crystalline film (SCF) as well as LuAP:Ce and YAlO$_3$:Ce (YAP:Ce) single crystals are reported. Room temperature absorption measurements of the single crystals in the vacuum UV region allowed establishing the bandgap energies of 7.63 eV for YAP and 7.86 eV for LuAP, with an assumption of the direct band-gaps. Luminescence of Ce$^{3+}$ in LuAP and YAP bulk crystals was measured as a function of temperature from 6 K up to 873 K. Temperature quenching of the Ce$^{3+}$ luminescence in YAP:Ce was observed above 650 K, which is related to the location of the lowest Ce$^{3+}$ *5d* level at 1.27 eV below the conduction band minimum. No temperature quenching occurred in LuAP:Ce up to 873 K, mostly due to the lower energy of the *4f* levels with respect to the valence band maximum. The barycenter energies and splittings of Ce$^{3+}$ *5d* states in YAP and LuAP at room temperature were precisely established. Theoretical calculations of the Ce$^{3+}$ *5d* states energy structure under pressure revealed a discrepancy between the obtained experimental results and the prediction of Dorenbos' theoretical model. The discrepancy can be removed if instead of the *5d* state of the free $Ce^{3+}$ ion the bandgap of the compound is taken as reference energy for the red-shift of the *5d* level. This hypothesis also allows us to take into account the pressure-induced increase of the bandgap energy, expected for the studied compounds. Pressure dependences of LuAP:Ce luminescence spectra suggest that a certain type of phase transition occurs above 15 GPa.

## *1. Introduction*

Yttrium and lutetium aluminum oxide perovskites ($YAlO_3$ – YAP, and $LuAlO_3$ – LuAP, respectively) are known to be prospective optical materials and have been studied for a relatively long time. Scintillators based on these materials are probably the most prominent application exploiting LuAP properties. Particularly, cerium-doped LuAP has a high light yield, high crystal density, and short decay time, which was already known back in 1998 [1]. Materials with high crystal density are the most suitable for scintillator applications, due to efficient γ-ray absorption. Therefore, LuAP seems to be more promising than $YAlO_3$:Ce, despite the same crystal structure and similar properties. Both materials could be also used for solid-state laser applications. They are also studied for a better understanding of their fundamental properties [2, 3].

LuAP was much less studied than the YAP, mainly due to difficulties in the growth process, related to non-congruent melting under higher temperatures than the YAP. To understand the mechanisms of LuAP:Ce luminescence better, it is important to know the location of the $Ce^{3+}$ states in the band gap of the host. Theoretical investigations of the $Ce^{3+}$ levels in the YAP band gap were performed in [2], but those results were not yet confirmed experimentally.

Both YAP and LuAP can be produced by many different growth methods, such as Czochralski, sol-gel, solid state reaction, liquid phase epitaxy, and others. Some materials grown by liquid phase epitaxy (LPE) contain less defects [4, 5]. In the case of LuAP, the LPE method is easier to apply than Czochralski since it requires much lower temperatures [6, 7, 8]. Using LPE method, single crystalline films (SCF) of high structural quality can be grown, with thicknesses from a few to a few tens of microns. Samples of such dimensions can be used in X-ray micro-imaging applications, where thin films with effective X-Ray absorption and high light yield are required [8, 9].

Despite applications and a relatively long period of studies, some basic properties of YAP and LuAP are surprisingly not well known. For example, the bandgaps of these materials were not measured with proper accuracy. The accurate bandgap measurements sometimes can be a difficult task since the band-gap energy depends on temperature and may differ by several hundreds of meV between cryogenics and room temperature. Also, some close-band-gap states, such as various types of excitons, Urbach states, and the presence of non-intentional impurities may hinder the position of the band-gap absorption. In addition, for a relatively large band-gap compound measurements with the use of vacuum-ultraviolet equipment of synchrotron radiation are required. Some procedures of band-gap estimation can give slightly different results. In this paper, we establish the values of the bandgaps of YAP and LuAP with the use of direct absorption measurements at room temperature. A relatively thin samples (a few tens of micrometer thick) were examined to get access to the band-gap absorption region.

Previous band-gap estimations of YAP and LuAP, among them also with the use of synchrotron radiation, produced slightly larger values of this parameter [10,11,12,13].The difference is in the order of 0.5 eV and it may be related to a different method of band-gap estimation used in the past and different temperatures of measurements.

Concerning doping with $Ce^{3+}$ some estimations of parameters of the energetic structure of this dopant in relation to the band structure of YAP and LuAP were done by Dorenbos [14]. However, the Dorenbos' estimations, although allowing having quite good qualitative information, sometimes differ by a few tenth parts of eV from those observed experimentally. This difference may be critically important for the evaluation of the thermal stability of luminescence efficiency of $Ce^{3+}$ dopant in particular compounds. These materials were not studied under high pressures, which affect their optical properties. This paper is dedicated to obtaining some general observations about the influence of high pressures on the energetic structure of $Ce^{3+}$ in various materials.

## 2. *Sample preparation and experimental techniques*

LuAP:Ce bulk crystals studied were grown by Czochralski and micro-pulling-down (μ–PD) methods. The Ce concentration was equal to 0.2 mol. %, 0.5 mol. % in YAP and 0.15 mol% in LuAP single crystals. $Ce^{3+}$ doped LuAP single crystalline films were grown by LPE method on (001) oriented YAP substrates using a melt solution containing $Lu_2O_3$ (5N), $CeO_2$ (5N), and $Al_2O_3$ (4N) and a PbO–$B_2O_3$ (5N) flux. The growth temperature was in the 1020–1035 °C range. The $CeO_2$ content in the melt-solution was 20 mol %. However, microanalysis of the films has shown that the Ce concentration was only equal to 0.055 and 0.05 mol % at 1020 °C and 1035 °C growth temperatures, respectively. Thus, the estimated segregation coefficient of Ce ions in LuAP SCFs lies in the 0.0025-0.005 range.

The measurements of single crystalline film surface morphology (see Fig. 1a) were performed using a JEOL JSM-820 scanning electron microscope (SEM), equipped with an EDS microanalyzer with IXRF 500 and LN2 Eumex detectors. The concentration of Pb (from the flux) and Pt (from the crucible) impurities in the layers were estimated to be in the 50-60 ppm range, respectively, i.e., below the 100 ppm level reported in [15]. The optically and structurally perfect SCF sample, with about 22 μm thickness, was chosen for high-pressure investigations of the $Ce^{3+}$ luminescence.

The single crystallinity and high quality of LuAP:Ce SCFs were confirmed by scanning electron microscopy (SEM) and X-ray diffraction (XRD) using a modified DRON 4 spectrometer (Cu $K\alpha_1$ X-ray source). The crystalline phase of the prepared LuAP:Ce/YAP epitaxial structures was characterized in the 2θ range from 20° to 100° with a step of 0.02°. The SEM image of the SCF surface presented in Fig. 1 (a) shows very few pits, hills, square holes, scratches, or brighter spots. The XRD ω-2θ scan in Fig. 1 (b) shows the (004) reflexes of the LuAP film and YAP substrate. From the difference in the peak positions, the misfit between the lattice constants of the YAP substrate and the LuAP SCF can be determined. The misfit, m = [($a_{scf}$ - $a_{sub}$)/$a_{scf}$] × 100%, is equal to -1.255% (see also Refs. 8, 16,17)

The large difference between the lattice constants of LuAP:Ce SCF and the YAP substrate causes huge mechanical stress on the film-substrate interface and can influence the optical properties of the LuAP:Ce film [18, 19]. The relaxation of the stress between the film and substrate may lead even to the cracking of the SCF sample. The propagation of such stress within the volume of SCF can cause gradual changes in the cation-anion distances and result in a notable change of the optical spectra as well of the LuAP host as of the dopants. Recently, a gradual change of the peak positions in the confocal Raman spectra and features of rare-earth luminescence was observed in LuAG:Ce SCFs grown by LPE on top of YAG substrate [20] as well as in SCF of TbAP perovskite grown onto YAP substrate[21]. The misfits between the respective garnet lattices and perovskite layers were about 0.73% and 1.3%, respectively [22].

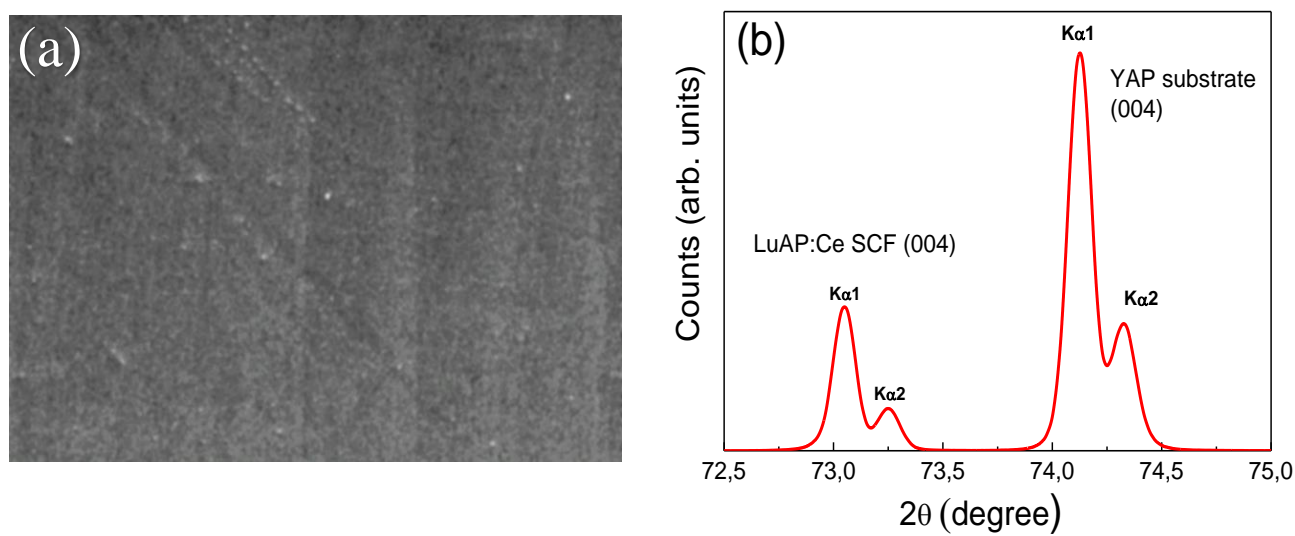

*Fig. 1. (a) SEM image of the LuAP:Ce SCF surface under 700 magnification. (b) XRD pattern of the (004) reflexes of the LuAP:Ce SCF film and the YAP substrate.*

The optical absorbance spectra were measured using a spectrophotometer JASCO V-660 with a double monochromator (1.5-6.5 eV) and a laboratory setup based on a vacuum monochromator VMR-2 and a hydrogen discharge light source (5.5-11 eV). In the latter case, the constant number of exciting photons was achieved by varying the slit width of the monochromator and using the constant signal from sodium salicylate for normalization.

An easyLab diamond anvil cell (DAC) was used for high-pressure (HP) measurements of LuAP:Ce SCF. Before the measurements, the samples were polished from the substrate side to a thickness of about 20-25 µm. Ruby crystals were used as pressure sensors. As pressure transmitting media either a 5:1 methanol-ethanol mixture or argon were applied in luminescence and Raman measurements, respectively. The applied pressure is hydrostatic up to about 10 GPa.[23] At higher pressures, it becomes quasi- hydrostatic, which means that the are some axial components of the applied pressure. It is possible to monitor the deviation from hydrostaticity by controlling the width of the luminescence lines of the ruby pressure sensor (typically it is done this way). In our case, the FWHM of ruby lines increased above 10 GPa but they were still well resolved. This means that non-hydrostaticity effects are not high.

Temperature dependences of the luminescence were measured in two different experimental setups. For measurements above room temperature, an FTIR 600 high temperature table provided by Linkam was used, while below room temperature an Oxford Optistat CF104 cryostat was used. The temperature overlapping region of both setups (from 77 K to room temperature) served to adjust the results. During both pressure and temperature-dependent measurements the same excitation source and registration system were used. Luminescence was collected in back-scattering geometry using a Triax 320 monochromator provided by ISA Yobin Yvon-Spex equipped with a Spectrum One liquid nitrogen cooled CCD camera. An Inova 400 argon-ion laser of constant power was used for 275 nm and 300 nm excitations.

Room temperature, unpolarized Raman spectra of LuAP:Ce SCF were collected with the use of a MonovistaCRS+ spectrometer equipped with: a 0.75 m Acton-Princeton monochromator; back-thinned, deep-depleted PyLoN system, liquid-nitrogen cooled CCD camera (1340×100 – 20 µm per pixel array); computer controlled Olympus XYZ IX71 inverted stage equipped with a Moticam (1280×1204) camera. The spectra were excited with the 532 nm laser line. The laser was focused on the sample through long working distance objectives with 50x (numerical aperture NA = 0.9) and 10x (NA = 0.23) magnification for experiments at ambient and high pressures, respectively.

The calculations of the structural and electronic properties of the LuAlO₃ crystal were performed using the CASTEP [24] program within the general gradient (GGA) and local density approximations (LDA), respectively [25]. The electronic configurations for Lu, Al, and O were taken as $4f^{14}5p^65d^16s^2$, $3s^23p^1$ and $2s^22p^4$, respectively. The convergence parameters were: for energy -5×10⁻⁶ eV/atom, maximum force 0.01 eV/Å, maximum stress 0.02 GPa, and maximum displacement 5×10⁻⁴ Å. The cut-off energy used for electronic properties calculations was 390 eV and the k-points grid was 9 × 7 × 10, which ensured the separation of the k-points in the reciprocal space of at least 0.02 1/Å. The initial LuAlO₃ structural data for the geometry optimization were taken from [26]. The calculations were performed in the range of pressures from 0 to 30 GPa with a step of 5 GPa. Although the calculated band gap is underestimated (which is typical for DFT calculations), the pressure coefficient of the calculated band gap correlates well with the experimental data.

### 3. Experimental results

### 3.1. Absorption spectra

Room temperature absorption spectra of Ce doped YAP and LuAP single crystals are shown in Fig. 2 (a) and (b), respectively. In the insets the square of the absorption coefficients of undoped YAP and thin Ce doped LuAP single crystals in the spectral region of 5–8 eV are presented, together with linear fits to the data near the fundamental absorption edges. The fits yield precise values of the energy gaps of YAP (7.63 eV) and LuAP (7.86 eV) at room temperature, assuming that the bandgaps are direct. Fits with an assumption of indirect band-gap are worse and yield slightly smaller bandgaps (for about 0.2-0.3 eV) [27], however, they agree very well with the bandgap energies established from the intrinsic luminescence excited by synchrotron radiation [28]. The determined bandgap energy of the LuAP bulk crystal is significantly lower than the previously reported value of 8.44 eV [29] determined from reflectivity for single crystalline films as well as the estimated energy of 8.2 eV for bulk LuAP [30].

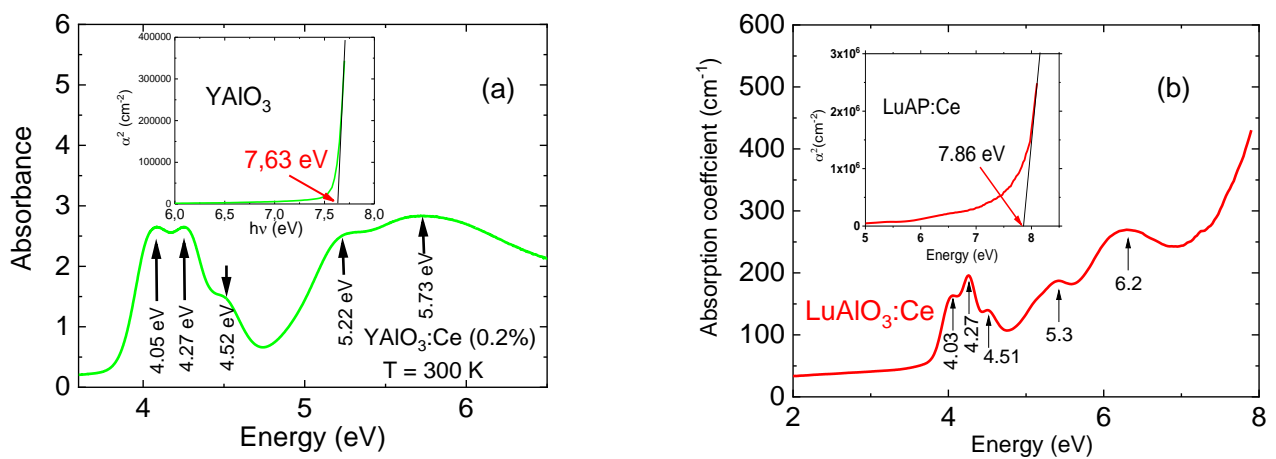

*Fig. 2. Room temperature absorption spectra of YAP:Ce (a) and LuAP:Ce (b) single crystals. The positions of Ce³⁺ 5d levels in LuAP:Ce and YAP:Ce crystals are marked with arrows. In the insets linear fits to the square of the absorption coefficients near the fundamental absorption edges of YAP (a) and LuAP (b) are shown.*

### 3.2. Temperature dependence of luminescence spectra

Luminescence of Ce$^{3+}$ in bulk YAP and LuAP single crystals was measured under 300 nm excitation as a function of temperature. As can be seen in Fig. 3 the Ce$^{3+}$ luminescence spectra in both samples are very similar. At low temperatures, they are composed of two broad bands corresponding to transitions from the lowest *d*-level to the spin-orbit split $^2F_{5/2}$ ground and $^2F_{7/2}$ excited states of the *4f* configuration (Fig. 3 (a)). Relatively small crystal field splitting of the two *4f* levels is not observed in the *5d→4f* luminescence due to strong electron-phonon coupling of the *5d* levels and low transition probability to the highest-energy component of the $^2F_{7/2}$ excited state [31].

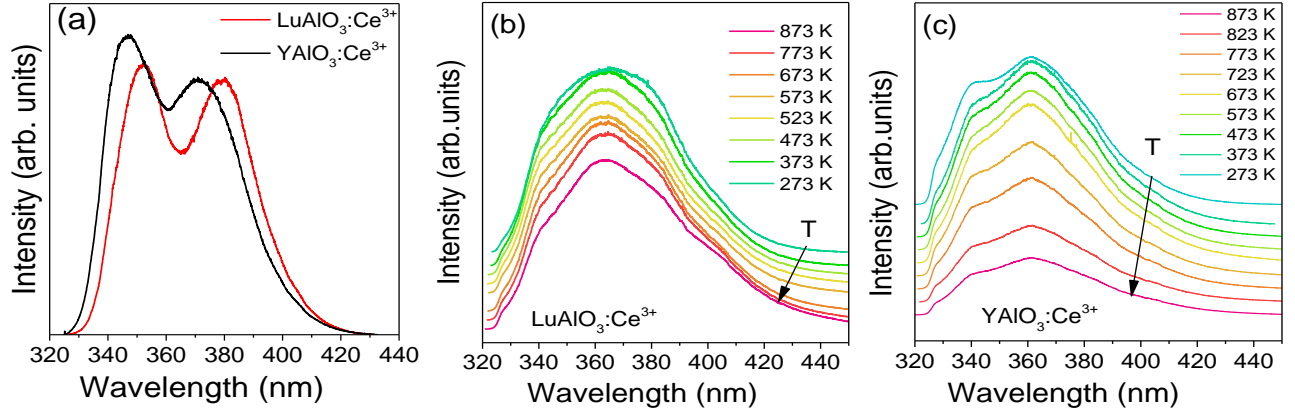

*Fig. 3. (a) Luminescence spectra of Ce$^{3+}$ in LuAP (red line) and YAP (black line) single crystals at 6 K. Temperature dependent luminescence spectra of Ce$^{3+}$ in LuAP (b) and YAP (c) single crystals. All spectra were detected under 300 nm excitation. The spectra in (b) and (c) are shifted in intensity scale.*

With increasing temperature, the second-lowest *d*-level, which lies at 237 nm and 216 nm in YAP:Ce and 234 and 200 nm in LuAP:Ce respectively (see Fig. 2a and 2b, respectively), (see Fig. 2), becomes thermally populated. Radiative transitions from this level lead to a slight blue shift of the emission spectra and to apparent smearing out of the double peak structure. Similar results were observed in [32]. The spectra collected at selected temperatures above 273 K in LuAP and YAP single crystals are shown in Fig. 3 (b) and (c), respectively.

As can be seen in Fig. 3 (b) the luminescence intensity in the LuAP:Ce crystal does not change with temperature up to 873 K, while in YAP quenching of Ce luminescence takes place already above 650 K (Fig. 3 (c)). The integrated Ce luminescence intensities normalized to room temperature ones are shown in Fig. 4 for YAP (full symbols) and LuAP (open symbols).

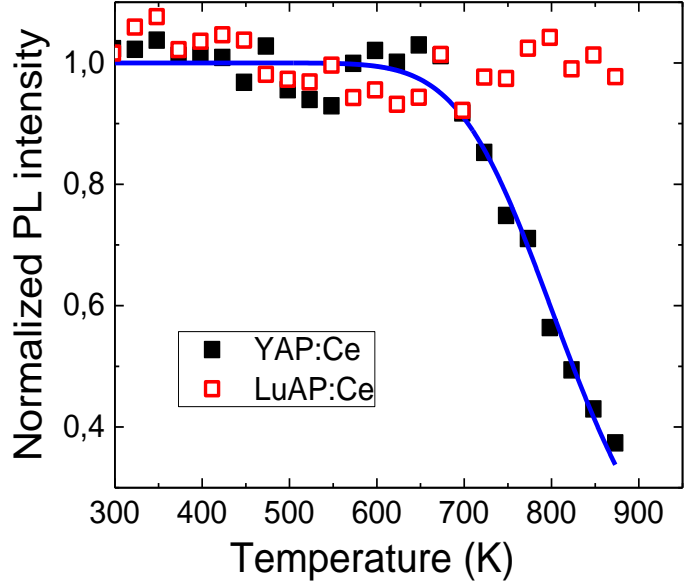

*Fig. 4. Temperature dependence of the normalized, integrated intensity of Ce³⁺ luminescence in LuAP (opened symbols) and YAP (full symbols). The solid line is a fit to Eq. (1).*

In the analysis of YAP:Ce temperature quenching the luminescence intensity was taken to be proportional to the population of the three lowest excited states multiplied by the corresponding radiative transition probabilities. Higher lying *5d* levels were neglected since the broadening of the absorption peaks indicates that they already degenerate with the conduction band. Equation (1) describes the luminescence intensity as a function of temperature, assuming that the electrons which are thermally activated to the second and third excited levels of Ce³⁺ contribute to the luminescence, while the electrons which are thermally excited to the conduction band are not recaptured and thus do not contribute to the emission. This assumption is valid given the existence of numerous electron traps responsible for thermoluminescence. Equation (1) takes thus the form:

$$I(T) = I_0 \frac{1 + a_1 e^{-\frac{E_{21}}{kT}} + a_2 e^{-\frac{E_{31}}{kT}}}{1 + a_1 e^{-\frac{E_{21}}{kT}} + a_2 e^{-\frac{E_{31}}{kT}} + a_3 e^{-\frac{\Delta E}{kT}}} \quad (1)$$

where $I_0$ is the initial low temperature luminescence intensity, $\Delta E$ is the energy distance between the lowest *5d* level and the bottom of the conduction band (treated as a fitting parameter), $E_{21}$ is the energy distance between the lowest *5d* and second lowest *5d* level, and $E_{31}$ is the energy distance between the lowest *5d* and third *5d* level. The parameters $\alpha_1$ and $\alpha_2$ are ratios of the radiative recombination probabilities from the second and third d-level to that of the lowest d-level, while $\alpha_3$ is the ratio between the ionization rate and the radiative transition probability. The $E_{21} = 0.22$ eV and $E_{31} = 0.470$ eV energies were estimated from the absorption spectra in Fig. 2 (a). From the obtained fit the lowest excited level of Ce³⁺ is located at $\Delta E = 1.27$ eV under the bottom of the conduction band. The parameters of the fit are given in Supplementary Information. The fit of the equation (1) to the experimental data is presented by a solid line in Fig. 4.

The obtained data allow estimating the position of the ground *4f* state of Ce$^{3+}$ in YAP crystal. This subject is very important since it is one of the basic parameters in the Dorenbos model [33]. Recent theoretical DFT calculations supported the value of 3.5 eV as the position of the first *4f* level of Ce$^{3+}$ above the top of the valence state [2]. Our results (i.e. distance between the lowest *5d* level and the bottom of the conduction band, equal to 1.27 eV, the bandgap energy equal to 7.63 eV, and the energy of the absorption peak to the first *5d* level – 4.05 eV) locate the ground *4f* level at 2.31 eV above the top of the valence band of YAP. This gives approximately 1.19 eV difference between the experimentally estimated position and the one obtained from Dorenbos theory and theoretical DTF calculations [2]. However, taking into account that the position of the first *4f* level is at the energy of the zero-phonon line (not observed experimentally due to very large electron-phonon coupling), located in the center between the absorption and luminescence bands maxima, one obtains a value of 3.76 eV for the distance between the lowest *4f* and *5d* levels. This decreases the difference by about 0.29 eV, locating the lowest *4f* level at 2.6 eV above the top of the valence band. The temperature decrease of the band-gap energy can be at least partly responsible for the remaining difference of 0.9 eV, yet not accounted for [34], assuming that the temperature change of the band-gap energy is related mainly to the change of the position of the bottom of the conduction band. Although the temperature dependence of the band-gap energy of YAP has not been experimentally established, the expected changes of the band-gap energy between room temperature and 873 K are in the range of at least a few hundreds of meV [34]. Therefore the theoretical predictions of the Dorenbos model and DFT calculations are not far from our experimental findings, within the expected accuracy limits (about 0.3 – 0.5 eV) [35]. We note, however, that the theoretical energy value of 3.5 eV must be overestimated since Ce$^{3+}$ luminescence quenching would then occur already at cryogenic temperatures.

The temperature quenching of the Ce$^{3+}$ luminescence is not observed in LuAP up to 873 K. It may be partly related to the larger bandgap of LuAP than YAP, although the relatively small difference between them (only 0.23 eV) indicates that the position of the ground state of Ce$^{3+}$ in LuAP must be closer to the top of the valence band than in YAP.

### 3.2. Raman spectra of the single LuAP:Ce crystalline film

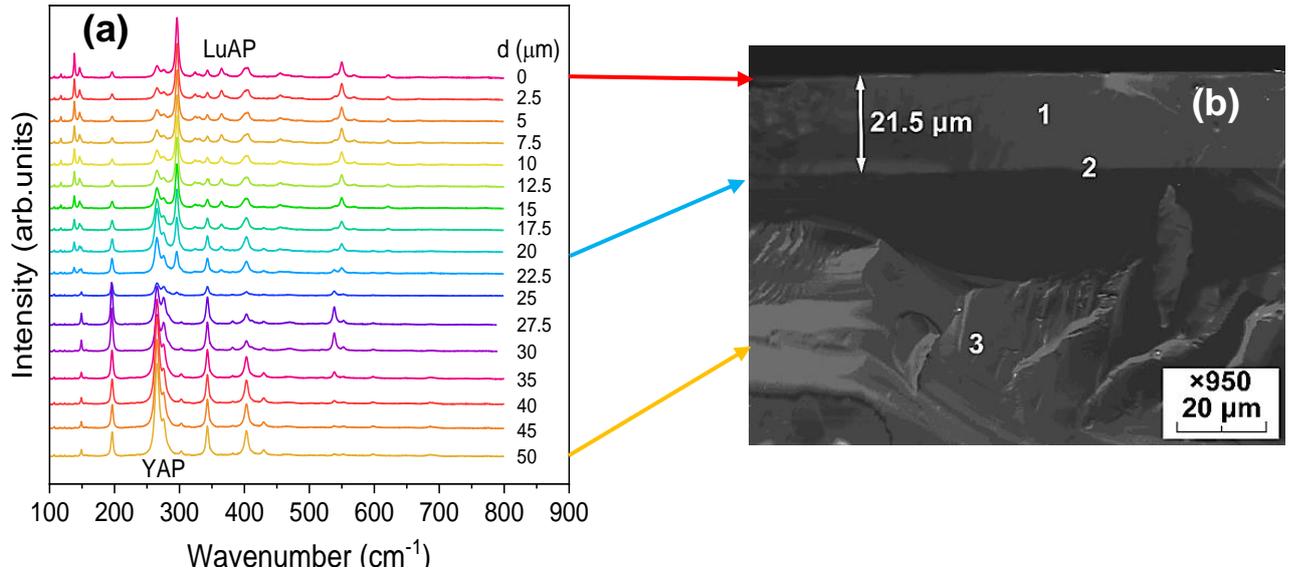

*Fig. 5. (a) Raman spectra of the LuAP:Ce single crystalline layer with thickness of about 22 μm grown on YAP substrates scanned along the YAP/LuAP cross section. The distance of the laser focus from the top of the SCF layer for each spectrum is given in the legend. (b) SEM picture of the YAP/LuAP cross section.*

Raman spectra of LuAP:Ce SCF with a thickness of approximately 21.5 microns grown on YAP substrate are shown in Fig. 5 (a). The measurements were performed on a cross-section of the LuAP/YAP layer/substrate, the SEM image of which is shown in Fig. 5 (b). The distance of the position of the laser spot from the top of the LuAP SCF layer, d, is given in Fig. 5 (a). The numbers 1, 2, and 3 in Fig. 5 (b) indicate the LuAP SCF layer, the LuAP/YAP interface, and the YAP substrate, respectively.

A list of the observed Raman lines together with their assignations are given in Table S1 in Supplementary Information. The Raman spectra of the YAP substrate resemble very much of the data reported in [18]. We observe altogether 20 Raman peaks for the YAP substrate (from the 36 predicted by group theory). Within the LuAP layer, 28 modes are observed, although only 11 of them occur solely in LuAP. Changes in the Raman spectra with the exciting beam moving across the YAP/LuAP structure are observed when the beam crosses the border between the YAP substrate and LuAP layer grown on it. This occurs at a distance of about 22 microns from the top of the LuAP layer. The structure of the Raman lines registered for LuAP SCF is similar to that observed for YAP. Additional theoretical studies are needed to assign the observed spectrum to vibrational modes expected for LuAP.

### 3.3. High-pressure luminescence

Luminescence of the LuAlO₃:Ce³⁺ bulk crystal and the LuAlO₃:Ce³⁺ single crystalline film (SCF) measured as a function of pressure under excitation with the 275 nm laser line is shown in Fig. 6 (a) and (b), respectively. The SCF sample loaded to the DAC was an almost free-standing

LuAP:Ce SCF layer since the thickness of the sample in the cell cannot exceed 25 microns. As can be seen in Fig. 6, the luminescence spectra in both samples are very similar. They are asymmetric and composed of at least two Gaussians. Also, the behavior under pressure is similar. The luminescence peaks shift towards longer wavelengths with increasing pressure and their intensity decreases. The intensity decrease is due to the movement of the *5d* levels under pressure while the wavelength of the excitation light remained constant.

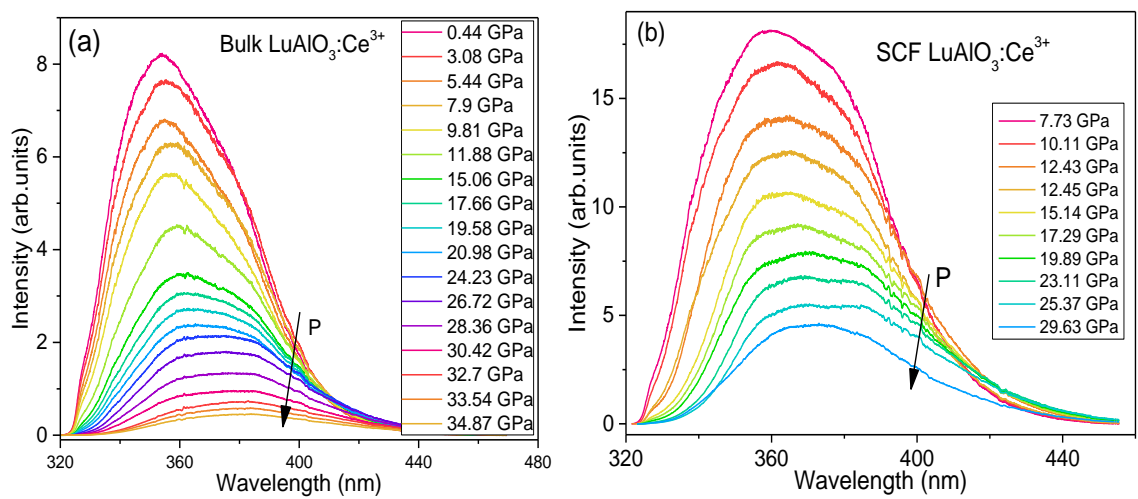

*Fig. 6. Room temperature luminescence spectra of $Ce^{3+}$ in $LuAlO_3$ single crystal (SC) (a) and single crystalline film (SCF) (b) under high pressure. The luminescence was excited with the 275 nm laser line and a 325 nm edge filter was used to cut off the laser excitation.*

All spectra were fitted with two Gaussians. The peak positions as a function of pressure are shown in Fig. 7 (a) and (b) for SCF and SC samples, respectively. Since the *4f* electrons are shielded from the influence of the crystal field by the filled outer shells, no significant changes in the level positions under pressure are expected. On the contrary, the *5d* electrons are much affected by the crystal field. Therefore, pressure-induced changes of the luminescence peak energies reflect predominantly the changes of the *5d* level position (from which the emission originates), while the energy distance between the two emission lines remains almost constant.

As shown in Fig. 7 (a) and (b), both luminescence lines shift towards lower energies with increasing pressure. The pressure dependences in the SCF sample exhibit a certain bending above 15 GPa. The red shift of the peak energies in the bulk crystal can be approximated by two linear dependences with shift rates of 16 and 19 cm$^{-1}$/GPa below 20 GPa and 47 and 54 cm$^{-1}$/GPa, above 20 GPa. The relative intensities of the higher to lower energy luminescence peaks are presented in Fig. 7 (c) and (d) for SCF and SC, respectively. These ratios decrease linearly with pressure up to about 15 GPa in the film and 20 GPa in the single crystal. At higher pressures, the dependences change noticeably. This behavior correlates perfectly with that of the peak positions. Certainly, the emission comes from not only the lowest energy 5d band but also from the higher energy components of excited 5d states due to their thermal population, even at room temperature.

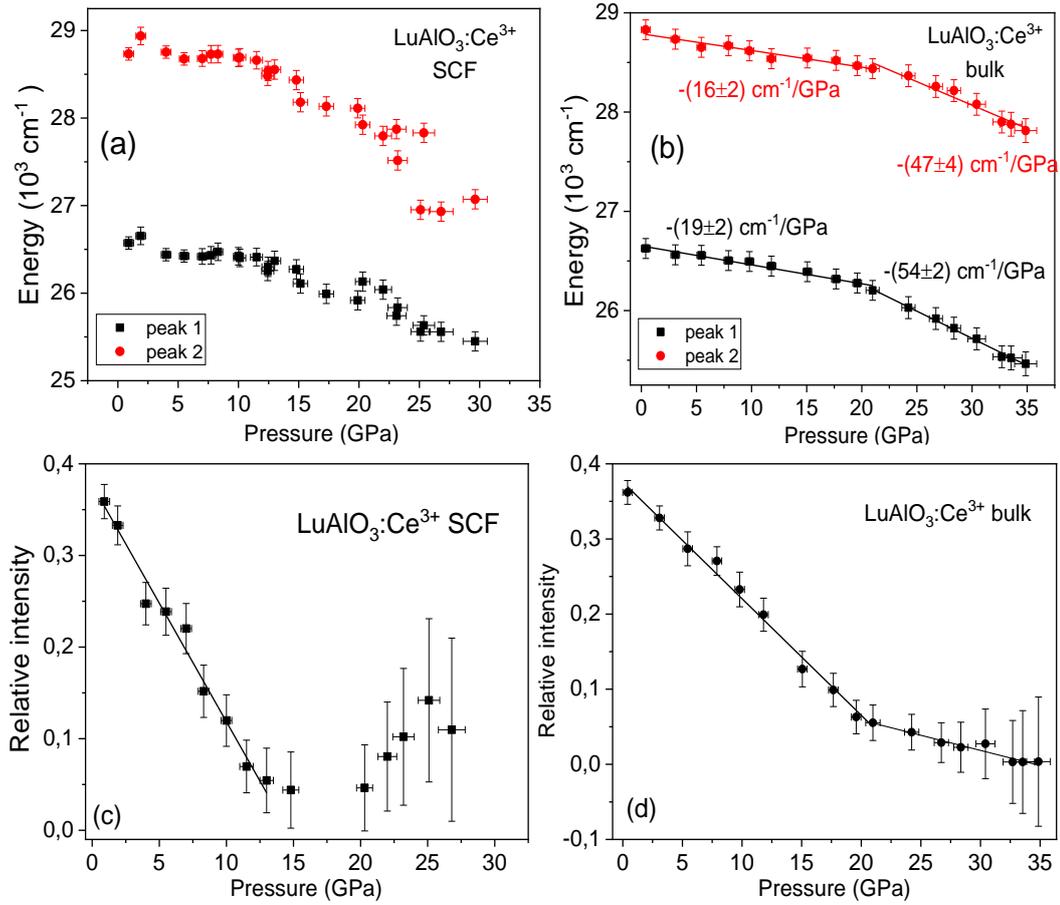

*Fig. 7. Energies (a,b) and relative intensities(c,d) of Ce³⁺ luminescence peaks associated with the 5d→ ²F₅/₂ and 5d→²F₇/₂ transitions vs. pressure in SFC and SC samples, respectively. The linear fits of the slopes are given in (b).*

The change in the rates is unexpected and suggests some structural changes taking place in both samples. One of the possible explanations is a phase transition occurring in this pressure range. In the case of the SCF sample, the pressure dependences of the peak energies are not quite parallel (see Fig. 7a). This suggests some influence of pressure on the *4f* electronic states. Indeed, a certain change in the splitting of the *4f* states may occur, as well as changes in the distance between the lowest-lying *5d* levels. The latter could lead to population changes of the *5d* states. In such a case, the luminescence would be initiated from more than just one level, especially at higher temperatures. Therefore, decomposition of the luminescence spectrum only into two bands is only an approximation. However, attempts to deconvolute the luminescence into more than two bands increase the number of fitting parameters, which effectively makes such a procedure unreliable.

The different pressures at which the change of shift rates is observed, i.e. around 15 GPa in SCF and around 20 GPa in SC samples, suggest that the bulk crystal is more pressure resistant. This can be related to the larger number of defects, which seem to play a structure stabilizing role. The lower defect concentration in the film than in the bulk crystal can be due to the much lower growth temperature. The different pressures at which the change of shift rates is observed, i.e. around 15 GPa in SCF and around 20 GPa in SC samples, suggest that the bulk crystal is more pressure resistant. This can be related to the larger number of defects, which seem to play a structure stabilizing role. The lower defect concentration in the film than in the bulk crystal can be due to the much lower growth temperature. Indeed, LuAP SCFs are prone to stress relaxation, associated with the difference in lattice parameters between the substrate and the overgrown layer, which sometimes even results in breaking.

### 3.4. Raman spectra under pressure

Pressure-dependent Raman spectra of the LuAP:Ce SCF single crystalline film are presented in Fig. 8 (a). The particular lines were grouped according to their similar pressure dependences and marked on the graph with numbers from 0.1 up to 6.4. At low pressures, the lines below 150 cm$^{-1}$ related to pressure-transmitting Ar are very well visible in the spectra, however, it is still possible to distinguish the low energy vibrations of the free-standing LuAP. With the increase of pressure, most of the lines observed on the cross-section of the YAP/LuAP structure are still visible. Above 10 GPa, however, some of the lines disappear, such as those from groups 1 and 6. Almost all lines exhibit shifts towards higher energies. The rate of the shift changes at the pressure of about 10 GPa, which coincides with the behavior of luminescence. The pressure dependences of Raman lines energies are shown in Fig. 8 (b). The shift rates are listed in Table S1 in Supporting Information for pressures below 10 GPa and above 15 GPa, respectively.

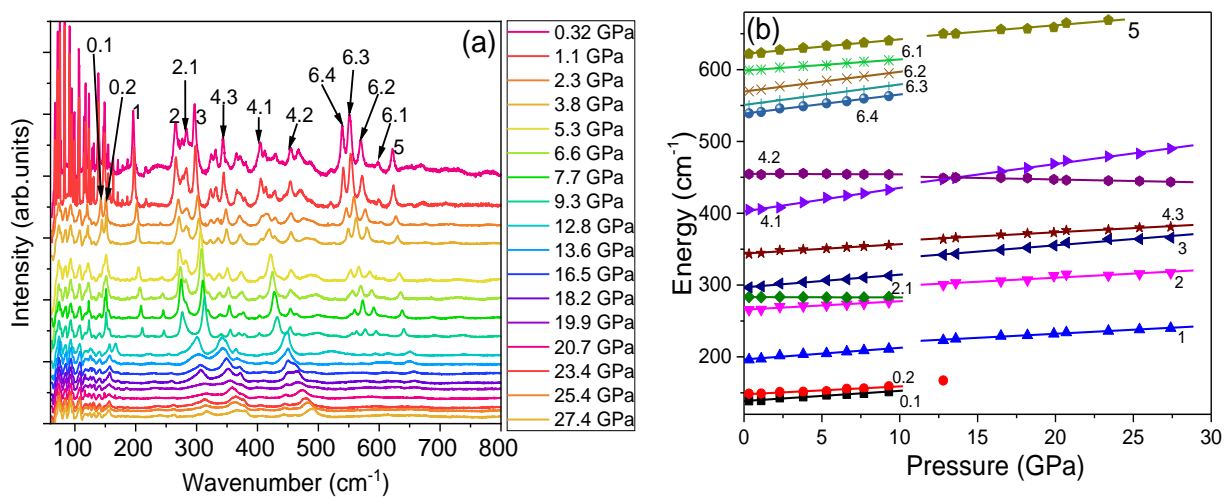

*Fig. 8. Pressure dependent Raman spectra of LuAP SCF. (b) Positions of Raman peaks as a function of pressure.*

An unusual exemption from the typical behavior, i.e. increase of the energy of the Raman modes with applied pressure, exhibits line 4.2, which energy decrease with applied pressure. This behavior can be seen in Fig. 9 (a). The shift rates for this line also change at a pressure of about 10 GPa, from -0.07 cm⁻¹/GPa below 10 GPa to -0.47 cm⁻¹/GPa above 15 GPa. This type of behavior resembles that of the so-called soft mode, observed, for example, in ferroelectric materials. Further increase of pressure in such materials leads to a ferroelectric/paraelectric structural phase transition [36]. This also suggests that a phase transition may be responsible for the observed shift changes in pressure coefficients of the luminescence bands and Raman lines.

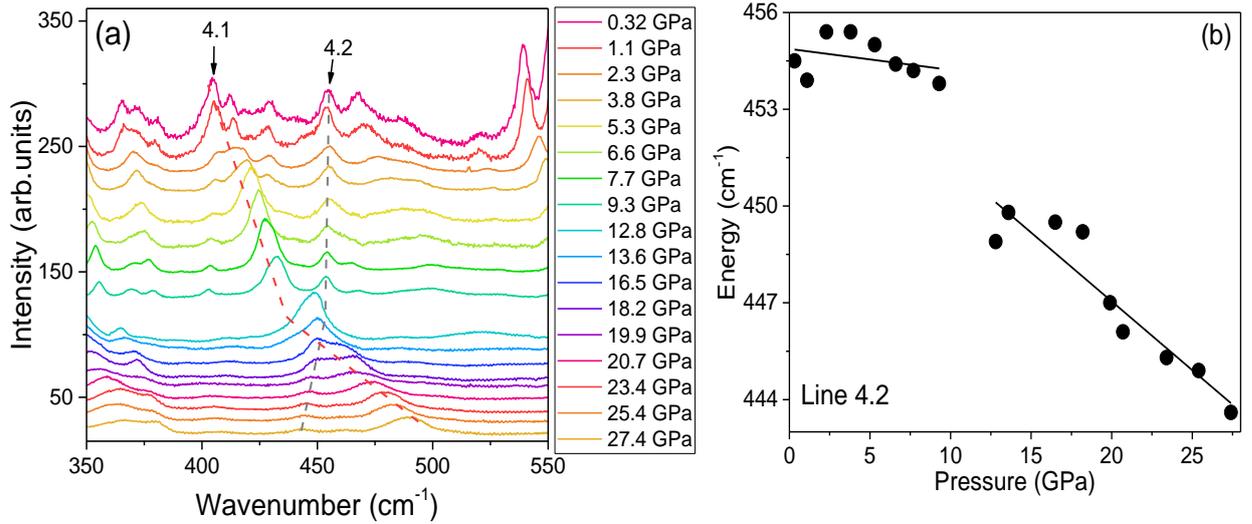

Fig. 9. Raman peak with diminishing energy for LuAP SCF (a) and its energy as a function of pressure (b)

However, another interpretation of this effect is also possible. Namely, the energy of certain vibrations in perovskite lattice may be dependent not only on the lattice parameters, which increase the energy with an increase of pressure, but also on an angle of deviations of the octahedra, forming the lattice, which may cause the changes of the force constants between the vibrating species [26]. Due to such distortions, the force parameters may decrease, which causes the energies of certain lattice vibrations may undergo a decrease with increasing pressure. The vibrations around 400 cm⁻¹ are associated with Al-O stretching and deformation modes in the parent YAlO₃ compound [37]. These effects do not exclude that an abrupt change of the already mentioned angle of distortion may occur at a certain pressure.

## 4. Theoretical calculation of the band-gap energy as a function of pressure

The results of theoretical calculations performed for LuAP with the use of two approximations: GGA and LDA, are presented in Fig. 10. The relative changes of unit cell volume under pressure are shown in Fig. 10 (a). Using calculated data the bulk modulus and its pressure derivative were calculated from the fit to Murnaghan equation of state [38]. Both approximations give similar values of these parameters (shown in the graph). They are in good agreement with the data already published for YAP [39].

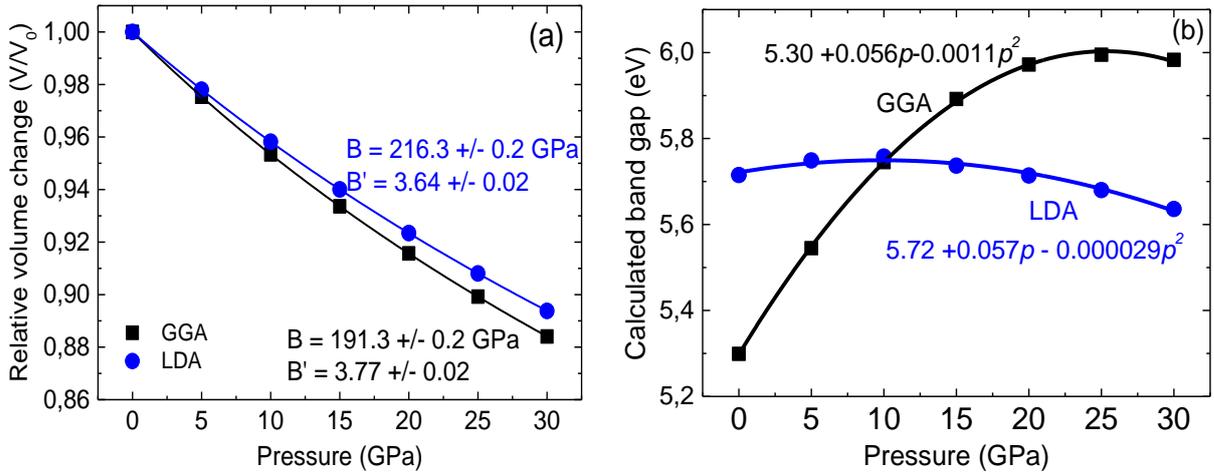

*Fig. 10. Calculated pressure dependences of the relative unit cell volume (a) and bandgap energy (b) of LuAP using GGA and LGA approximations. Fitted values of bulk moduli and their pressure derivatives are given in (a) for both approximations. The bandgap pressure dependences can be fitted with quadratic functions presented in the graph.*

The calculated pressure dependence of LuAP bandgap energy is shown in Fig. 10 (b), again for both used approximations. As typical for these calculations, the value of the bandgap energy is underestimated, although the pressure changes of this parameter are usually in better agreement with experimental data [40, 41]. However, as can be seen, GG and LD approximations give quite different pressure dependences of the LuAP bandgap energy. This is most probably related to relatively close energies of the direct and indirect bandgaps, similar to the energy structure of YAP [2]. Unfortunately, it is impossible to measure directly this dependence due to the absorption of diamonds, which prevents performing such an experiment. Results of GGA predict that the LuAP bandgap increases strongly with applied pressure, but at higher pressures, a very strong bowing is observed when the indirect bandgap prevails over the direct one. This effect is observed at pressures higher than 15 – 20 GPa. From LDA calculations a very weak influence of pressure on the band-gap energy is expected. Pressure dependences of the bandgap energy derived from both theoretical methods can be approximated by quadratic functions, shown in Fig. 10 (b) with appropriate parameters.

### 5. Discussion

The redshift of $Ce^{3+}$ luminescence observed in the high-pressure experiments is related to the compression and possible distortion of the crystal lattice. Physical mechanisms leading to the red shift are three-fold: (i) down-shift of the barycenter (centroid) of *5d* levels, $\varepsilon_c$, as compared to the free ion, (ii) increase of the crystal field splitting of the 5d states, $\varepsilon_{cfs}$, (iii) distortion of the oxygen polyhedron surrounding $Ce^{3+}$ from cubic symmetry. The first two mechanisms can be evaluated in the following way: the down-shift of the centroid can be described by the following formula, adopted (modified) from papers [42, 43, 44]:

$$\varepsilon_c = A \sum_{i=1}^{N} \left[ \frac{\alpha_{sp}^i}{(R_i - 0.6\Delta R)^6} \right] \qquad (2)$$

where $R_i$ is the distance (pm) between $Ce^{3+}$ and anion $i$ in the undistorted lattice. Pressure application changes these distances. The summation is over all N (N = 8) anions that coordinate $Ce^{3+}$. 0.6 $\Delta R$ is a correction for lattice relaxation around $Ce^{3+}$, $\Delta R$ is the difference between the radii of $Ce^{3+}$ and $Lu^{3+}$, $\alpha_{sp}^i$ (in units $10^{-30}$ $m^{-3}$) is the spectroscopic polarizability of anion $i$, and A is a constant ($1.79 \times 10^{13}$). The polarizability $\alpha_{sp}^i$ can be evaluated from the formula:

$$\alpha_{sp}^i = 0.33 + 4.8/\chi_{av}^2 \qquad (3)$$

where $\chi_{av}$ is the weighted average of the electronegativities of the cations in the oxide compounds.

The contribution of the splitting of the *5d* state by the crystal field can be evaluated from the empirical formula:

$$\varepsilon_{cfs} = \beta_{poly} R_{av}^{-2} \qquad (4)$$

where $R_{av}$ is the average distance between the activator and the neighboring anions, $\beta_{poly}$ values are in the ratio 1, 0.89, 0.79, 0.42 for octahedral, cubic, dodecahedral, and tricapped trigonal prismatic coordination, respectively, and $\beta_{octahedral} = 1.35 \times 10^9$ $pm^2$ /cm [45].

Crystal field splitting of the *5d* levels is also dependent on the distortion from the cubic geometry of the octahedron, which is more difficult to evaluate.

Results of the calculations, based on the Dorenbos theory [43] and ambient pressure data, with the use of distances between the cation and surrounding oxygens taken from [46] for YAP and from [26] for LuAP, respectively, are presented in Fig. 11. The distances $R_i$ are listed in Supplementary Information, together with other data necessary to calculate formulas (2), (3), and (4). In the calculations, we assumed that Y, Lu, and consequently, Ce ions are located in dodecahedral (however strongly distorted) crystallographic positions, thus the values of $\beta_{dod} = 0.79 \times \beta_{oct}$. The experimental positions of the barycenters of *5d* states of $Ce^{3+}$ ions, estimated from the integral of absorption measurements of all *5d* transitions for each compound (see Fig. 2), are also given in Table 1, together with the barycenter energies obtained as the mean energy of the absorption peaks. The experimental values are compared to theoretical ones, obtained by subtracting the calculated centroid shift, $\varepsilon_c$, from the *5d* barycenter for the free $Ce^{3+}$ ion, i.e., 6.35 eV (51 230 $cm^{-1}$) [44]. The crystal-field splitting, $\varepsilon_{cfs}$, is defined as the energy difference between the lowest and highest *5d* level. A fraction of $\varepsilon_{cfs}$ contributes to the total down shift. Our estimations show that this fraction is equal to about 0.65 at ambient pressure and it was kept constant in the calculation of pressure dependences. A schematic picture of the energy structure of Ce³⁺ ions in YAP and LuAP under ambient and high pressure is presented in Fig. 11.

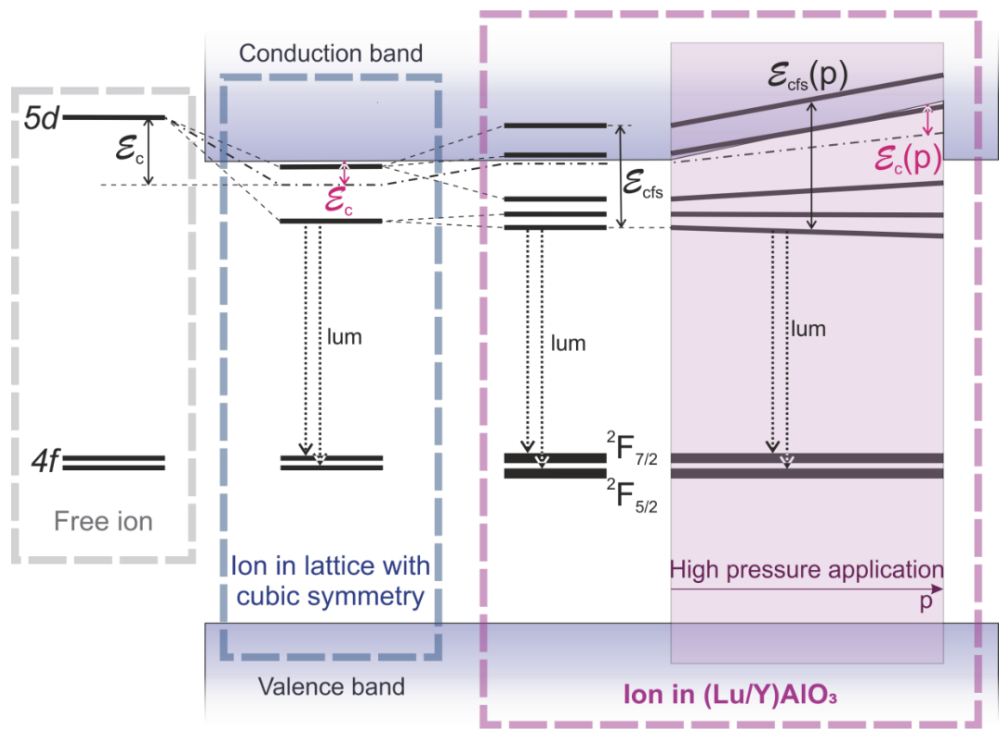

*Fig. 11. Schematic energy structure of Ce³⁺ ions in YAP and LuAP hosts under high pressure.*

*Table 1. Calculated from formulas (2), (3), and (4), and experimental values of the down-shift of the centroid barycenter, ε_c, the splitting of the 5d state by the crystal field, ε_cfs, and related data at ambient pressure (see description below).*

| Parameter | YAP:Ce | | LuAP:Ce | |
|---|---|---|---|---|
| | theory (eV) | experiment (eV) | theory (eV) | experiment (eV) |
| Bandgap energy | | 7.63 | | 7.86 |
| $\varepsilon_c$ | 2,58 | | 3.06 | |
| $\varepsilon_{cfs}$ | 2,30 | 1.68 | 2.38 | 2.17 |
| Theoretical barycenter energy $E^{5d}_{free}-\varepsilon_c$ | 3.77 | | 3.29 | |
| Experimental barycenter energy (vs. 4f state energy) | | (4.76)* | | (4.86)* |
| Theoretical ($\varepsilon_c+0.65\times\varepsilon_{cfs}$) and experimental ($\varepsilon_c-0.65\times\varepsilon_{cfs}$) down-shift** | $\varepsilon_c+0.65\times\varepsilon_{cfs}$ 4.08 | $\varepsilon_c - 0.65\times\varepsilon_{cfs}$ (3.67)* | $\varepsilon_c+0.65\times\varepsilon_{cfs}$ 4.61 | $\varepsilon_c - 0.65\times\varepsilon_{cfs}$ (3.56)* |
| Energy of lowest 5d level $E^{5d}_{free}$ − down shift | **2.28** | **4.05** | **1.74** | **4.03** |

| Energy of lowest 5d level referred to the band-gap | 3.55 | 4.05 | 3.24 | 4.03 |
|---|---|---|---|---|

*Calculated as an average of the level peaks energies.

** The difference in sign of $\varepsilon_{cfs}$ comes from different reference levels

The calculated theoretical energies of the lowest *5d* $Ce^{3+}$ states in YAP and LuAP referred to free-ion energies do not agree with the values obtained from absorption measurements, as shown in Table 1. The theoretical values are about twice smaller than those observed experimentally. A better agreement is achieved for the total splitting energy of the *5d* states, $\varepsilon_{cfs}$.

The *5d* barycenter energy of $Ce^{3+}$ in LuAP was estimated as equal to 5.22 eV if calculated from integrated absorption to the *5d* states. This value is much larger than obtained as the average of the observed peak energies of the five *5d* states of $Ce^{3+}$, which is equal to 4.86 eV. The value obtained from the integration of the absorption coefficient is affected by the relatively large background present in the LuAP crystal close to the conduction band. The *5d* barycenter in LuAP estimated from the integrating procedure is located between the two highest-lying *5d* levels. In contrast, in YAP the *5d* barycenter energies of $Ce^{3+}$ calculated both as an average of the peak energies and from the integration of the absorption coefficients agree with each other (see Table 1). The better agreement in YAP is a consequence of the much lower background absorption than in LuAP. Therefore, the experimental $Ce^{3+}$ *5d* barycenter energy in LuAP is better taken as equal to 4.86 eV, from the average position of *5d* levels peak energies.

The theoretical energies of the *5d* level positions much better correlate with the experiment if they are calculated from the bandgap energies instead of the free $Ce^{3+}$ ion *5d* state energy. The results of such calculations are shown in the last row of Table 1. Nevertheless, the difference of about 0.8 eV is still present in the case of LuAP:$Ce^{3+}$. The energy of the bandgap is very close to the energy of the $Ce^{3+}$ to $Ce^{4+}$ charge-transfer process [47].

Now it is possible to calculate the pressure dependence of the energies of *5d – 4f* luminescence bands in LuAP:Ce, assuming that all distances between $Ce^{3+}$ ions and the surrounding ligands change according to the modified Murnaghan equation of state [38] (assuming, that the distances under pressure change as $V^{1/3}$):

$$\frac{R_0}{R_p} = \left(\frac{pB_0'}{B_0} + 1\right)^{\frac{1}{3B_0'}} \qquad (5)$$

where $R_0$ and $R$ are $Ce^{3+}$ – oxygen distances at ambient and applied pressures, respectively, $B_0$ and $B'_0$ are values of bulk modulus and its pressure derivative [48]. Fixing the theoretical positions of the *5d – 4f* band maxima to the experimental positions at ambient pressure, the theoretical pressure dependences of the luminescence bands were calculated. They are shown in Fig. 12 as broken lines. The apparent discrepancy between experimental and theoretical data can be corrected by taking as a reference level the position of the conduction band bottom instead of the free-ion *5d* level energy. The best fit to the experimental data is obtained if a linear shift of the conduction band minimum with pressure, equal to 310 cm$^{-1}$/GPa, is assumed (see solid blue lines in Fig. 12).

Taking the pressure dependence of the bandgap calculated with the use of LDA (red lines in Fig. 12) leads to results very similar to those obtained for a fixed reference level, which is not in agreement with the experimental data. The dependence calculated with the use of theoretical GGA predictions (green solid lines) gives good agreement for pressures up to about 15 GPa, but at higher pressures, the theoretical lines are bent off experimental data.

There is no experimental data on the pressure dependence of the bandgap energy of LuAP due to its high energy, coinciding with the bandgap of diamonds. However, both the linear dependence of the band gap energy and this calculated with GGA are in reasonable agreement with typically observed in similar compounds, for example YGG [34]. Therefore, we postulate the use of the bandgap energy as the reference level instead of the free Ce ion *5d* energy. This small correction allows the correlation of experimental data with the Dorenbos model. It is in line with his idea of Vacuum Referred Binding Energy (VRBE) or Host Referred Binding Energy (HRBE) concept, relatively recently presented [49, 50].

The obtained results, with linear dependence of the bandgap energy or that calculated with GGA now exhibit good agreement between theory and experimental data, at least up to 15 GPa. They also show the importance of considering the pressure dependence of the band structure. There is no possibility to neglect pressure dependence of the energy of the conduction band minimum vs. the position of the *4f* levels, which in this consideration are treated as pressure independent. This is in agreement with the commonly accepted idea that *4f* states, being screened by outer electrons, are very weakly dependent on the environment. Contrary to that, the *d* levels, especially also *5d*, strongly interact with surrounding anions and therefore their energy structure is strongly pressure dependent.

The use of conduction band energy as the reference level for the calculation of the down-shift energy $\varepsilon_c$ may be justified by the close location of the intervalence charge transfer state IVCT, which is formed by ionization of the Ce$^{3+}$ ion followed by trapping of the ionized electron. This state can be understood as a (Ce$^{4+}$ + e) exciton and its energy level should lie close to the bottom of the conduction band depending on the electron binding energy [51].

Although the obtained general agreement between such theoretical considerations and experimental data is quite good, the more close comparison reveals also important deviations, which are detected by our measurements. First, the bending of theoretical curves in Fig. 12 (for linear band-gap dependence) is opposite to the observed experimentally (compare also with Fig. 7a and Fig. 7b). This may be associated with experimental details not taken into account in the above considerations, i.e.: (i) the pressure dependence of the cation – anion distances may not be the same for all anions, as assumed in * (4); (ii) the local compressibility of the polyhedron around $Ce^{3+}$ may be different than the average described by Murnaghan's equation; (iii) the distortion from cubic symmetry may result in stronger splitting than described by Eq. (3); finally (iv) a certain type of phase transition may occur at pressures between 15 and 20 GPa. This possibility is suggested by the changes of the pressure coefficients of the Raman spectral lines, which also occur at similar pressures. Also, the evident deviation from experimental data of the theoretical pressure dependence of the $5d{\rightarrow}4f$ transitions for the band-gap calculated with GGA above 15 GPa may be associated with this phase transition. We would rather not expect a phase transition to a different crystallographic structure or different space group symmetry, however, there is a possibility that the angles, at which the octahedra containing Al ions are placed in the lattice, may undergo an abrupt change under pressures in the range 15 GPa – 20 GPa. This will not change the symmetry of the crystal lattice, however, may affect the rigidity of the crystal structure, allowing significant changes of the pressure coefficients of the phonon modes and luminescence peaks.

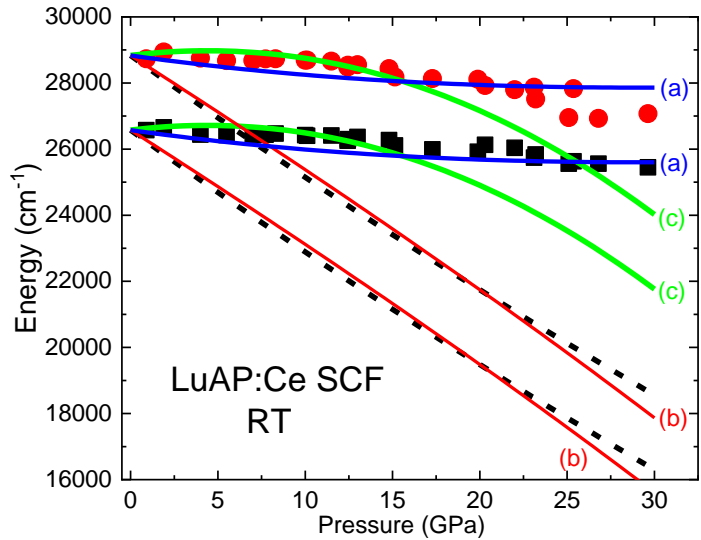

*Fig. 12. Pressure dependences of 5d→4f transitions calculated from Dorenbos theory. Broken lines show dependences calculated according to Eqs. (2), (3), and (4), solid lines present dependences calculated taking into account changes of the energy of conduction band minimum. Blue lines: linear pressure dependence of the band gap energy(a); red lines: band-gap calculated with LDA (b); green lines: bandp-gap calculated using GGA (c).*

A certain kind of phase transition is the most probable explanation for the observed change of pressure coefficients of the $Ce^{3+}$ $5d{\rightarrow}4f$ transitions at about 15 GPa in the free-standing layer and at a slightly higher pressure in the crystal grown by the micro-pulling down technique. A similar effect was observed in [52], associated with certain reconfiguration of octahedrons containing Al ions, however, without a change of the general structure, such as the compound's space group. The postulated changes are reversible due to this. We also observed a similar effect for REAP:$Eu^{3+}$ layers [53]. The different pressures at which this transition occurs in the free-standing LuAP layer grown by liquid phase epitaxy and in the crystals grown by micro-pulling down may be related to different types and concentrations of defects existing in both materials, especially in the LuAP crystals grown at high temperature (2000 $^{o}$C) than their SCF counterparts (~1000 $^{o}$C). Namely, the concentration of $Lu_{Al}$ antisite defect and oxygen vacancies and their aggregates are substantially higher in SC than in SCF [11]. Also, the concentration of uncontrollable RE impurities seems to be higher in SC than in SCF. A relatively large number of unidentified RE impurities was detected by luminescence measurements of SC.

The possibility of a phase transition is confirmed also by high-pressure Raman measurements depicted in Fig. 8. In this experiment, the change of pressure coefficients of various lines is also observed in a similar pressure range (above 10 GPa). The observation of a soft mode in the Raman spectra, which is a very characteristic feature of possible phase transitions, gives an additional hint that this happens in the studied material.

### 5. Conclusions

The results of absorption measurements in a near-UV region of $YAlO_3$ and $LuAlO_3$ crystals allowed accurate measurements of the bandgap energies of $YAlO_3$ and $LuAlO_3$ single crystals at room temperature, which are equal to 7.63 eV and 7.86 eV, respectively, assuming direct bandgaps. Thermal quenching of YAP:Ce luminescence observed above 650 K in high temperature measurements, locate the position of the lowest excited *5d* level of $Ce^{3+}$ at 1.27 eV below the bottom of the conduction band. This value is affected by the temperature change of the band-gap, i.e. at low temperatures, the difference between the energy of the *5d* level and the bottom of the conduction band is larger. The estimated position of the *4f* level is thus consistent, within the experimental error, with the estimations of Dorenbos theory and DFT calculations.

$Ce^{3+}$ luminescence quenching is not observed in LuAP crystals up to about 873 K. This is a result of the larger bandgap of LuAP as compared to YAP as well as the lower energy of the ground $Ce^{3+}$ *4f* states, which are located closer to the top of the valence band in LuAP than in YAP.

The down shift of the *5d* energy levels of $Ce^{3+}$ with respect to the energy of the free $Ce^{3+}$ ion calculated according to Dorenbos theory does not agree with experimental data. The difference can be reconciled if the down shift is calculated relative to the bandgap energy of YAP and LuAP. This approach also allows correlating the observed changes of the *5d* state energies under pressure in LuAP, related to the pressure-induced changes of the average cation-anion distances, assuming that the main changes of the bandgap are due to the energy increase of the conduction band bottom.

The other possibility proposed in [54] is a pressure-induced shift of the energies of both *4f* and *5d* manifolds. This effect would lead to a lack or very little dependence of the *4f*↔*5d* transition energies on pressure. This is, however, in contradiction to the Dorenbos model and the common expectation that *5d* states are more influenced by the ligands than *4f* ones.

We suggest that the observed change of the pressure coefficients of the *5d→4f* $Ce^{3+}$ luminescence bands is associated with pressure-induced structural transitions, occurring in the liquid-phase epitaxy grown layers at a pressure of about 15 GPa, and at higher pressures crystals grown by the micro-pulling down grown method. We related this difference to the larger number of unintentional impurities and structural defects present in the micro-pulled down crystals than in the single crystalline film.

The observed abrupt changes of the pressure coefficients of Raman modes above 10 GPa in LuAP confirm the hypothesis of a phase transition. In the high pressure Raman experiment also one soft mode with an energy equal to 455.4 $cm^{-1}$ at ambient pressure was identified, the energy of which decreases with pressure.

*Acknowledgements:*

This work was partially supported by the Polish National Science Centre projects no. 2018/31/B/ST8/03390 and SENG-2 2021/40/Q/ST5/00336.

M.G. Brik also thanks the supports from the Chongqing Recruitment Program for 100 Overseas Innovative Talents (Grant No. 2015013), the Program for the Foreign Experts (Grant No. W2017011) and Wenfeng High-end Talents Project (Grant No. W2016-01) offered by Chongqing University of Posts and Telecommunications (CQUPT), Estonian Research Council grant PUT PRG111, European Regional Development Fund (TK141) and Polish NCN project 2018/31/B/ST4/00924.